\documentclass[%
reprint,
twocolumn,
superscriptaddress,
amsmath,amssymb,
aps,
floatfix,
]{revtex4-2}
\usepackage{amsfonts}
\usepackage{setspace}
\usepackage{dcolumn}
\usepackage{bm}
\usepackage{tikz}
\usepackage{ulem}
\usepackage{graphicx}
\usepackage[colorlinks=true,citecolor=blue,urlcolor=blue]{hyperref}
\usepackage{amsmath,amsfonts,amssymb,times,natbib}

\newtheorem{thm}{Theorem}
\newtheorem{cor}{Corollary}

\begin{document}

\title{Conservation Laws, Dynamical Back-Action, and the Non-Classicality of Gravity}
\title{Conservation Laws and the Non-Classicality of Gravity}

\author{Tianfeng Feng}
	\affiliation{ QICI Quantum Information and Computation Initiative, Department of Computer Science, The University of Hong Kong, Pokfulam Road, Hong Kong}

  \author{Chiara Marletto}	
  \affiliation{Clarendon Laboratory, University of Oxford, Parks Road,	Oxford OX1 3PU, United Kingdom}
\author{Vlatko Vedral}
	\affiliation{Clarendon Laboratory, University of Oxford, Parks Road,		Oxford OX1 3PU, United Kingdom}

\begin{abstract}
We analyze the interaction between quantum matter and classical objects through a general effective channel for hybrid dynamics, subject to the fundamental constraint that no quantum correlations can be generated between the classical and quantum sectors from any initially separable state. We demonstrate that, within this hybrid framework, imposing an additive conserved observable $\langle O_{QC} \rangle = \langle O_Q \rangle + \langle O_C \rangle$ strictly forbids a classical system from altering the local observable $\langle O_Q \rangle$ of its quantum counterpart. Applying this no-go theorem to gravity, under the assumption of such hybrid dynamics, we show that if global momentum or energy is conserved, a strictly classical gravitational field cannot induce momentum or energy transfers in a quantum system. In contrast, a quantum gravitational field naturally facilitates such back-action. Drawing upon the fundamental relationship between conservation laws and the quantum properties of objects, our analysis provides a novel perspective for interpreting existing experimental observations, such as free fall, as potential indicators of the non-classicality of gravity.

\end{abstract}
	

	\maketitle

The quantization of gravity remains one of the most profound and elusive quests in modern theoretical physics, seeking to reconcile the continuous spacetime geometry of general relativity with the discrete, probabilistic nature of quantum mechanics. Despite significant theoretical endeavors—ranging from the canonical quantization of fields and the geometrization of quantum mechanics to the overarching framework of string theory \cite{DeWitt,Kibble,string}—a unified theoretical framework remains stubbornly out of reach. These enduring conceptual and mathematical challenges \cite{Kiefer,Isham} have prompted alternative propositions, including the hypothesis that gravity is fundamentally classical \cite{Carlip}, an emergent macroscopic phenomenon \cite{Penrose,Sakharov}, or simply not a fundamental force requiring quantization at all \cite{Shanarin}. Consequently, the necessity of quantizing the gravitational field has become a subject of intense and ongoing debate \cite{Eppley, Huggett,Albers,Kent,Garcia,CMVVwhy}, exacerbated by the extreme difficulty of obtaining direct empirical evidence at the Planck scale \cite{Dyson,Coradeschi}.

As Feynman perceptively noted during the 1950 Chapel Hill conference, the physical state of a composite system exhibits fundamentally distinct properties depending on whether a mass becomes quantumly entangled with the gravitational field or remains merely classically correlated with it \cite{Feynman}. This insight underscores the critical role of macroscopic spatial superpositions in probing the quantum regime of gravity. Recently, Feynman's foundational idea has experienced a significant revival. Novel proposals suggest that observing gravitationally induced entanglement between two mesoscopic masses would provide an unambiguous operational witness for the non-classicality of the gravitational mediator \cite{Bose,CMVV2017}.  While conceptually elegant and theoretically robust \cite{Plenio,Tianfeng,Rijavec,Kim}, with several advanced experimental protocols now formulated \cite{Krisnanda,Howl,Carney,MP,Bose2,Datta}, the exceptionally weak nature of gravitational coupling means that realizing these entanglement-based tests remains a formidable technological challenge for the foreseeable future.

Given these experimental hurdles, a natural question arises: can existing, accessible empirical observations offer insights into the non-classical nature of gravity? In this paper, we propose a rigorous framework to address this possibility. There have also been some gedankenexperiment demonstrating the  paradoxes of causality and complementarity brought about by classical gravity \cite{Belenchia,Danielson}. 
Moving beyond pure thought experiments, recent advancements in quantum information theory have demonstrated that imposing fundamental physical principles—such as macroscopic conservation laws or operational uncertainty relations—can effectively constrain the nature of a mediator, even without direct entanglement verification \cite{LG,CM,Tajima,CMVV2}. 

Historically, extensive literature has explored so-called quantum-classical hybrid dynamics. However, the systems in these traditional frameworks are often quasi-classical, inherently retaining underlying quantum algebraic structures \cite{hd,Diosi2014,koopman,Sudarshan,Peres,Diosi2000,Hall}. Building upon information-theoretic approaches to physical interactions \cite{CM,Tajima}, we investigate the intrinsic relationship between macroscopic conservation laws and gravitational non-classicality by imposing a strict definition of classicality. We relax standard assumptions, such as the linearity of the overall dynamics. Instead, we model the hybrid interactions through a generalized operational framework: we require that any purely classical mediator must admit an effectively sequential decomposition into sector-local operations and classical-to-quantum control channels. This explicit structural restriction guarantees the absolute inability of the classical sector to generate quantum correlations.

Within this rigorous framework, we derive a fundamental no-go theorem: for any dynamics constrained by such an effectively sequential decomposition, it is impossible to dynamically alter the local observables of its subsystems while strictly preserving an additive global conserved quantity. Applying this limitation to gravitational interactions yields an important physical consequence.
This theoretical restriction poses a severe challenge to the compatibility between strictly classical gravitational channels and standard kinematic observations, such as the free fall of a massive particle. Conversely, our framework mathematically demonstrates that a gravitational field possessing non-classical degrees of freedom naturally accommodates such dynamical back-action. Consequently, within this specific hybrid operational regime, macroscopic conservation laws coupled with everyday kinematic observations may serve as an operational witness for the non-classicality of gravity.

\section{Definition of a  Classical system}

A fundamental distinction between classical and quantum systems lies in their contextuality. In this paper, we rigidly define a ``classical'' system as one characterized by a single, preferred observable (i.e., a single basis, such as $\{|i\rangle \langle i|\}$). All other dynamical variables of such a classical system are strictly functions of this fundamental observable \cite{CMVV2,CMVVwhy}. This definition not only enforces that all permissible physical states are perfectly distinguishable, but it also fundamentally constrains the system's interactions: any coupling to an external system must exclusively involve these mutually commuting observables \cite{Tanjung,CMVV2017,CMVV4}.

A direct corollary of this algebraic restriction is that such a classical system can only sustain classical correlations with any other entity, including quantum systems \cite{Modi}. Lacking non-commuting observables, the classical sector is mathematically incapable of generating quantum entanglement. Furthermore, this strict asymmetry dictates that while a classical system can coherently control operations on a quantum system, the reverse---quantum-controlled operations on the classical sector---is strictly forbidden, as it would inevitably induce non-classical superposition within the classical system \cite{Reasons}.

\section{ General Quantum-classical Dynamics }

Classical correlations manifest in the majority of physical dynamic processes. These correlations can arise between quantum-quantum systems, classical-classical systems, or quantum-classical systems. In the subsequent discussion, our attention will be directed towards exploring general quantum-classical hybrid systems. 
Before proceeding, we will provide the necessary notations and definitions for both quantum and classical systems.

 For a quantum system, we utilize the density matrix in Hilbert space to represent a quantum state. On the other hand, for a classical system, we do not make any assumptions about a specific model. 
 To maintain generality, we denote a pure classical state as  ``$\text{State}_i$'',  where each unique value of $i$ represents a perfectly distinguishable state. Naturally, for a mixed classical state, it can be defined as follows:
$ \wp_{\mathcal{C}}=\sum_{i} p_i \text{State}_i$,
where $p_i$  is the probability of different pure states ``$\text{State}_i$'' (for continuous classical system, $\wp_{\mathcal{C}}=\int_x p(x) \text{State}_x dx$). In the following, without loss of generality, we focus on the discrete case, as the continuous case can be approximated by the discrete case with arbitrarily large dimensions and arbitrary precision.

As classical systems cannot be quantum correlated with quantum systems, their composite states exhibit solely classical correlations. Suppose the initial state of the hybrid system lacks classical correlation,  following a quantum-classical interaction, the resulting state generally evolves and can be expressed in the following general form:
\begin{equation} \label{CC}
	\wp_{\mathcal{C}}\oplus \varrho  \rightarrow	\sum_{i} q_i \wp_{{\mathcal{C}}i} \oplus \varrho_i,
\end{equation}
where $\{q_i\}$ is the probability distribution of classical correlation and $\wp_{{\mathcal{C}}i}=\sum_{j}p_j \text{State}_i$.  Historically, the evolutionary state of a hybrid system is called the hybrid density operator \cite{hd,Diosi2014}, which satisfies an overall normalization condition and preserves the form of the state. 

\begin{figure}[tb]
	\center
	\includegraphics[scale=0.5]{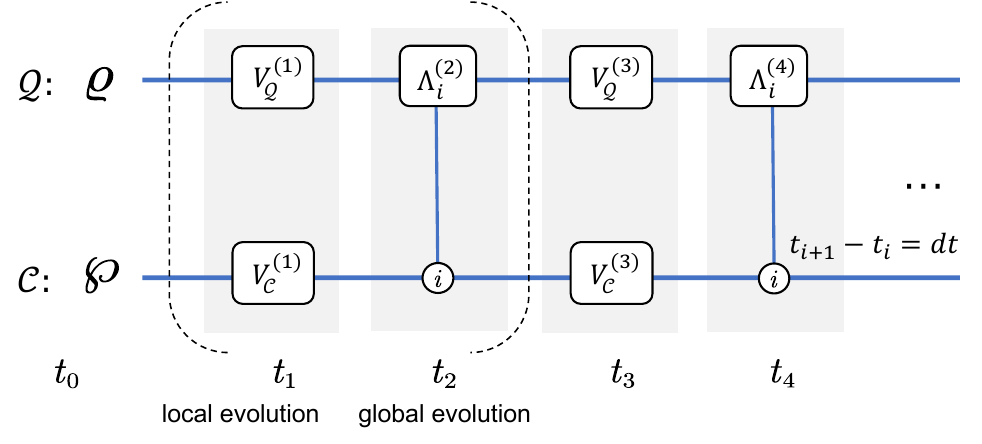}
	\caption{ 
Sequential decomposition of a general effective channel for hybrid quantum-classical dynamics. The continuous evolution is rigorously discretized into infinitesimal short-time slices, $dt = t_{i+1} - t_i$. Within each infinitesimal interval, the joint dynamics strictly factorize into local, non-entangling free evolutions on the respective subsystems, sequentially followed by a conditional classical-to-quantum control channel. This precise decomposition explicitly forbids the generation of quantum correlations.}
	\label{fig1}
\end{figure}

An operational definition of quantum-classical dynamics $\Theta_{\mathcal{QC}}$ can be derived based on the limitations of classical correlations and the properties of classical systems. As depicted in Figure \ref{fig1}, we provide a general channel for quantum-classical dynamics over infinitesimal time intervals. 
In this context, we refrain from imposing any specific theoretical framework on quantum-classical dynamics and instead focus on its effective quantum channel decomposition, emphasizing how it generates classical correlations. 

This decomposition aligns with the principles of physical causality. In non-relativistic quantum mechanics, the Hamiltonian dictates that the evolution of interacting particles can occur simultaneously, i.e. $e^{-i(\sum_j H_j+ \sum_k H^k_{int}) t}$. However, in reality, interactions between atoms are mediated by electromagnetic fields, propagating at the finite speed of light. Consequently, the actual interaction processes are sequential rather than simultaneous, akin to Trotterization, i.e. $(\prod_j e^{-iH_j \delta t} \prod_k e^{-iH^k_{int} \delta t})^{t/{\delta t}}$. 
Furthermore, in the linear version of the Di\'osi-Penrose model, the ultimate master equation is also given by a sequential evolution \cite{DP}. This consists of a free evolution governed by the stochastic master equation for monitoring the density of masses, followed by the gravitational potential acting over an infinitesimal amount of time.  
Thus, even if whole operations may appear consistent, the sequential decomposition and whole operations embody distinct physical meanings. The global viewpoint may lose the underlying physics. 
It's important to note that while we are examining the infinitesimal time evolution of hybrid systems, this approach possesses the generality for straightforward extension to continuous time dynamics.


 Without loss of generality, here we analyze the hybrid dynamics of an infinitesimal time slice, as in figure \ref{fig1}, while the proof of continuous time is put into the Appendix.
We suppose the initial state of the hybrid system is uncorrelated, denoted as $\chi_{\mathcal{QC}}=\wp_{\mathcal{C}}\oplus \varrho$, where $\wp_{\mathcal{C}}=\sum_{i}p_i \text{State}_i$.  Generally, as shown in Figure \ref{fig1} (single segment), one may divide the quantum-classical dynamics into two distinct parts. Firstly, there is a map that is physically and effectively equivalent to the local action $V=V_{\mathcal{C}}\oplus V_{\mathcal{Q}} $
 where $V_{\mathcal{Q/C}}$ operating on the quantum /classical state $\varrho_\mathcal{Q}/\wp_\mathcal{C}$. Consequently, $V$ transforms the hybrid state $\chi_{\mathcal{QC}}$ into $V (\wp_{\mathcal{C}}\oplus \varrho)=V_{\mathcal{C}}(\wp_{\mathcal{C}})\oplus V_{\mathcal{Q}} (\varrho) =\sum_{i}q_i \text{State}_i \oplus \varrho^\prime
$. 
The second part involves the establishment of a control channel $\Lambda_i$ from the classical sector to the quantum sector, from which the classical correlations of the states of the composite system originate. The specific process of the second part can be expressed as:
\begin{equation}
\sum_{i}q_i \text{State}_i \oplus \varrho^\prime \rightarrow \sum_{i}q_i \text{State}_i \oplus \varrho_i,
\end{equation}
 where $\varrho_i=\Lambda_i (\varrho)$. 
 Therefore, one has $\Theta_{\mathcal{QC}} ( \chi_{\mathcal{QC}})=  \chi_{\mathcal{QC}} ^\prime =\sum_{i}q_i \text{State}_i \oplus \varrho_i$.
  We emphasize that the maps $\Theta_{\mathcal{QC}}$, $V$, and $\Lambda_i$ are not required to be state-independent; they may generally be functions of $\chi_{\mathcal{QC}}$, accommodating nonlinear mappings. Regardless of the underlying physical nature of the hybrid dynamics, it must ultimately manifest this classical correlation structure.
  
  This operational definition of an effective hybrid channel rests on two general axioms: (1) It must continuously map a legal physical state to another legal physical state \cite{hd,Diosi2014}. (2) A quantum system is strictly prohibited from imposing a control operation onto a classical system, because conditioning a classical state on a quantum superposition would inevitably induce non-classical entanglement, directly violating the fundamental definition of classicality.
  
  An intuitive way to comprehend this decomposition is to map the classical state $\{\text{State}_i\}$ to a set of strictly orthogonal bases $\{|i\rangle\langle i|\}$. Because the classical system possesses only this single basis, any arbitrary interaction with the quantum system can be exhaustively broken down into local free operations and classical-to-quantum control channels \cite{NC}. For example, if linearity holds, a standard quantum interaction like $e^{-i|j\rangle \langle j| \otimes A_j}$ is exactly reproducible by our control operation $\Lambda$ utilizing unitary operators $\Lambda_i$. However, by relaxing the assumption of linearity and strictly forbidding quantum-to-classical control, the most general possible form of quantum-classical dynamics over infinitesimal intervals is uniquely defined by the cascading circuit decomposition depicted in Figure \ref{fig1}.

\subsection{Observables in Hybrid systems}
    Physical attributes, precisely quantified by the expectation values of observables, are fundamentally determined by the interplay between the system's state and its dynamical variables. We can succinctly formalize this mapping as:
	\begin{equation}
	\langle \text{Observable} \rangle	\equiv \{\text{State, Observable}\} .
	\end{equation}
This generalized notation successfully encompasses both classical and quantum frameworks. In quantum theory, the expectation value of an observable $O$ evaluates to $\{\varrho, O\} \equiv \text{Tr}(\varrho O)$, where $\varrho$ denotes the quantum density matrix. (Note that the bracket $\{\cdot, \cdot\}$ here denotes a generic evaluation map, not an anti-commutator or Poisson bracket) .Without loss of generality, we restrict our analysis to discrete classical systems for simplicity. For a classical mixture described by $\wp_{\mathcal{C}} = \sum_i p_i \text{State}_i$, the expectation value is linearly defined as $\{\wp_{\mathcal{C}}, O\} \equiv \sum_i p_i \{\text{State}_i, O\}$. Crucially, if $\text{State}_i$ represents an eigenstate with a definitive scalar value $O_i$, this mapping exactly recovers the standard statistical expectation: $\sum_i p_i O_i$. Extending this formalism to a hybrid quantum-classical system, the expectation value of a global observable is naturally evaluated over the hybrid state $\chi_{\mathcal{QC}}$ as $\langle O \rangle = \{\chi_{\mathcal{QC}}, O\}$. For strictly additive observables of the form $O = O_{\mathcal{C}} + O_{\mathcal{Q}}$, the expectation value linearly decomposes into local contributions:
$$\langle O_{\mathcal{C}} + O_{\mathcal{Q}} \rangle = \{\chi_{\mathcal{QC}}, O_{\mathcal{C}}\} + \{\chi_{\mathcal{QC}}, O_{\mathcal{Q}}\} \equiv \langle O_{\mathcal{C}} \rangle + \langle O_{\mathcal{Q}} \rangle.$$
One can see the Appendix for a comprehensive treatment of hybrid observables.

\subsection{A No-go theorem}

Here we present a no-go theorem demonstrating the impossibility of a general quantum-classical dynamics (satisfying the decomposition in Figure \ref{fig1}) causing a change in the local observable of either a classical or quantum subsystem, provided the global observable is conserved. In other words, the strict conservation law imposes fundamental limitations on the dynamic exchange of corresponding local observables between classical and quantum systems. The no-go theorem is formalized as follows: 

\begin{thm}
For any quantum-classical dynamics admitting an effectively sequential decomposition into sector-local operations and classical-to-quantum control channels (see Figure \ref{fig1}), the expectation value of any local observable $\langle O_{\mathcal{C}/\mathcal{Q}} \rangle$ remains strictly invariant, provided the dynamics globally conserves the additive quantity $O = O_{\mathcal{C}} + O_{\mathcal{Q}}$.
\label{t2}
\end{thm}

To rigorously derive this theorem, suppose there is a universal conservation law for the observable $O$, such as energy or momentum. For any isolated physical system ($\mathcal{C}$ or $\mathcal{Q}$) evolving freely without interaction, the local observable $O_{\mathcal{C}/\mathcal{Q}}$ must inherently remain unchanged. For a composite quantum-classical hybrid system, the universally conserved observable takes the strictly additive form $O=O_{\mathcal{C}}+ O_{\mathcal{Q}}$, satisfying
\begin{equation}\label{cl}
	\langle O_{\mathcal{C}}+O_{\mathcal{Q}}\rangle_{t}= \langle O_{\mathcal{C}}+ O_{\mathcal{Q}}\rangle_{t+\Delta t},
\end{equation}
where $\langle O_{\mathcal{C}}+ O_{\mathcal{Q}}\rangle=\{\chi_{\mathcal{QC}}, O_{\mathcal{C}}\}+\{\chi_{\mathcal{QC}}, O_{\mathcal{Q}}\}=\langle O_{\mathcal{C}}\rangle+\langle  O_{\mathcal{Q}} \rangle$. The conservation law dictates that Eq.~(\ref{cl}) must hold at all times. This implies that, as illustrated in Figure \ref{fig1}, every infinitesimal slice of the dynamical evolution must adhere to this global conservation. Consequently, any local free operation and the subsequent global control operation must independently comply with the conservation of the respective observables.

Without loss of generality, suppose the initial state of the hybrid system possesses no classical correlation, formulated as $\chi_{\mathcal{QC}}=\wp_{\mathcal{C}} \oplus \varrho$, where $\wp_{\mathcal{C}}=\sum_{i} p_i \text{State}_i$. Immediately, the initial expectation value of the global observable $O$ at $t_0$ is evaluated as:
\begin{equation}
 \langle O_{\mathcal{C}}+ O_{\mathcal{Q}}\rangle_{t_0}=\sum_{i}p_i \{\text{State}_i, O_{\mathcal{C}}\}+\text{Tr}(\varrho O_{\mathcal{Q}}).
\end{equation}
Following the local operation $V$, the hybrid state transforms into $V (\wp_{\mathcal{C}}\oplus \varrho)=V_{\mathcal{C}}(\wp_{\mathcal{C}})\oplus V_{\mathcal{Q}} (\varrho) =\sum_{i}q_i \text{State}_i \oplus \varrho^\prime$. The expectation value at $t_1$ becomes:
\begin{equation}
	\langle O_{\mathcal{C}}+ O_{\mathcal{Q}}\rangle_{t_1}=\sum_{i}q_i \{\text{State}_i, O_{\mathcal{C}}\}+\text{Tr}(\varrho^\prime O_{\mathcal{Q}}).
\end{equation}
Since $V_{\mathcal{C}}$ and $V_{\mathcal{Q}}$ represent local free evolutions governed by the conservation law, they must satisfy $\sum_{i}p_i \{\text{State}_i, O_{\mathcal{C}}\}=\sum_{i}q_i \{\text{State}_i, O_{\mathcal{C}}\}$ and $\text{Tr}(\varrho O_{\mathcal{Q}})=\text{Tr}(\varrho^\prime O_{\mathcal{Q}})$. 

Subsequently, the classical-to-quantum control operation $\Lambda$ maps the state from $\sum_{i}q_i \text{State}_i \oplus \varrho^\prime$ to $\sum_{i}q_i \text{State}_i \oplus \varrho_i$, yielding the expectation value at $t_2$:
\begin{equation}
	\langle O_{\mathcal{C}}+ O_{\mathcal{Q}}\rangle_{t_2}=\sum_{i}q_i \{\text{State}_i, O_{\mathcal{C}}\}+\sum_{i}q_i\text{Tr}(\varrho_i O_{\mathcal{Q}}),
\end{equation}
where $\varrho_i=\Lambda_i (\varrho^\prime)$. Because $O$ is strictly conserved throughout the entire evolution, we must have $\langle O_{\mathcal{C}}+ O_{\mathcal{Q}}\rangle_{t_0}=\langle O_{\mathcal{C}}+ O_{\mathcal{Q}}\rangle_{t_1}=\langle O_{\mathcal{C}}+ O_{\mathcal{Q}}\rangle_{t_2}$. Canceling the invariant classical expectations on both sides necessitates:
\begin{equation}
\langle O_{\mathcal{Q}}\rangle_{t_0} =\text{Tr}(\varrho O_{\mathcal{Q}})=\text{Tr}(\varrho^\prime O_{\mathcal{Q}})=\sum_{i}q_i\text{Tr}(\varrho_i O_{\mathcal{Q}}). 
\end{equation}
This confirms that $\langle O_{\mathcal{Q}}\rangle_{t_0}=\langle O_{\mathcal{Q}}\rangle_{t_1}=\langle O_{\mathcal{Q}}\rangle_{t_2}$. Iterating this sequential logic, it naturally follows that over any extended temporal evolution from $t_0$ to $t_n$, $\langle O_{\mathcal{Q}}\rangle_{t_0}=\langle O_{\mathcal{Q}}\rangle_{t_n}$ precisely holds (see Appendix). This rigorously establishes that a hybrid dynamical model decomposing into classical-quantum interactions cannot alter the local expectation $\langle O_{\mathcal{Q/C}}\rangle$ while fulfilling the global conservation law.

In stark contrast, the fully quantum interaction between two quantum systems intrinsically allows for the dynamical exchange of local observables. In quantum theory, the exact conservation of the global observable $O = O_{\mathcal{Q}_1} + O_{\mathcal{Q}_2}$ requires:
\begin{equation}
\langle [O_{\mathcal{Q}_1}+O_{\mathcal{Q}_2}, U_{\mathcal{Q}_{1}\mathcal{Q}_2}] \rangle=0,
\end{equation}
where $U_{\mathcal{Q}_{1}\mathcal{Q}_2}$ dictates the unitary evolution of the composite system. However, the local operators generally do not commute with the joint unitary evolution, yielding $\langle [O_{\mathcal{Q}_i}, U_{\mathcal{Q}_{1}\mathcal{Q}_2}] \rangle \ne 0$ for $i=1,2$. Therefore, a fully quantum mediator fundamentally liberates the local observable:
\begin{equation}
\langle O_{\mathcal{Q}}\rangle_{t} \ne \langle  O_{\mathcal{Q}}\rangle_{t+\Delta t}.
\end{equation}
Note that the no-go theorem derived above seamlessly extends to continuous classical systems described by $\wp_{\mathcal{C}}=\int_x p(x) \text{State}_x dx$.

Capitalizing on this profound operational distinction between hybrid and fully quantum dynamics, we immediately establish the following crucial corollary:

\begin{cor}
Suppose a quantum system $S$ interacts with an unknown mediating entity $E$ under the strict conservation of an additive global observable $\langle O_S \rangle + \langle O_E \rangle$. If this interaction dynamically alters the local observable $\langle O_S \rangle$, then the joint evolution cannot be governed by the sequential hybrid decomposition of Figure \ref{fig1}. Consequently, the mediator $E$ must fundamentally possess non-classical degrees of freedom.
\label{c2}
\end{cor}

That is to say, subject to an additive conservation law, if a mediator of unknown nature dynamically induces a transfer of a corresponding local observable in a quantum system, it serves as a robust operational witness that the mediator is fundamentally quantum-like, entirely ruling out this broad class of hybrid quantum-classical descriptions.

	\section{Quantumness of gravitational field}\label{QforGraivity}

\begin{figure}[tb]
	\center
	\includegraphics[scale=0.45]{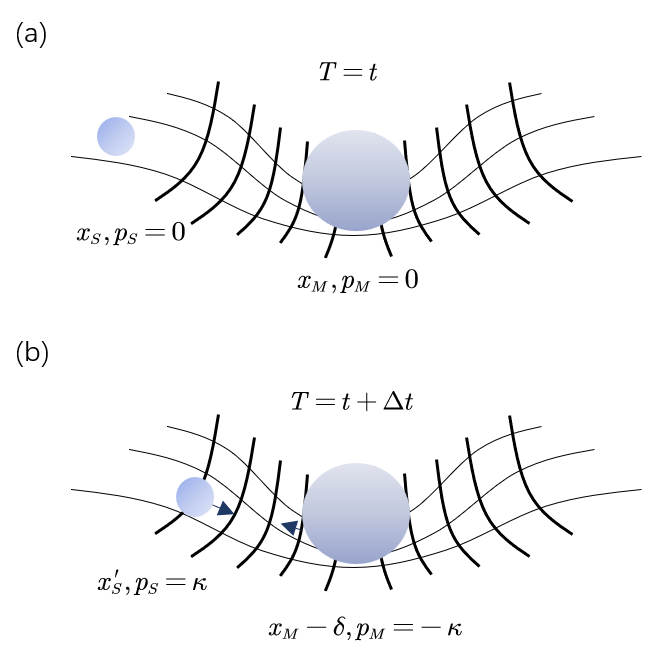}
	\caption{
    FIG. 2. Dynamical momentum transfer between two quantum masses mediated by the gravitational field $E$. (a) At time $T = t$, the two masses are stationary at initial positions $x_S$ and $x_M$, with zero initial momenta ($\langle P_S \rangle = \langle P_M \rangle = 0$). (b) At $T = t + \Delta t$, the masses have undergone spatial displacement. Their respective momenta acquire equal and opposite changes, strictly preserving the global momentum conservation of the composite system.
	}
	\label{fig2}
\end{figure}

We now leverage our established no-go theorem for hybrid quantum-classical channels to formally argue for the non-classical nature of the gravitational field. We premise our argument on the principle of strict locality: interactions between spatially separated objects are not instantaneous action-at-a-distance, but are explicitly mediated by a local field. Consequently, we elevate the gravitational field—sourced by a mass \cite{FG, CMVV2017, Galley2,Test2}—to the status of an independent dynamical entity, akin to the electromagnetic field. Our objective is to determine whether this mediating entity must inherently possess quantum degrees of freedom. For formal clarity, we model the Newtonian gravitational field as an explicit physical system, denoted as $E$. 

Consider a bipartite matter arrangement consisting of a quantum probe mass $S$ and a macroscopic source mass $M$, as illustrated in Figure \ref{fig2}(a). At time $T=t_0$, both masses are stationary. Empirical physics dictates that these masses will gravitationally attract; their interaction is locally mediated by the field $E$. While the global momentum of the hybrid composite $S \oplus E \oplus M$ system is strictly conserved at zero, the individual masses $S$ and $M$ acquire equal and opposite momentum dynamically, as shown in Figure \ref{fig2}(b). Crucially, by modeling the composite state via the direct sum $\oplus$ rather than the tensor product $\otimes$, we strictly forbid any quantum correlation between the classical field and the quantum masses. 

We now analyze this standard phenomenon under the stringent assumption that the mediating field $E$ is strictly classical. Given the locality of the interaction, the dynamics decomposes into local bipartite couplings $S-E$ and $E-M$, with no direct interaction between $S$ and $M$. Subjecting this arrangement to our no-go theorem (\autoref{c2}), and enforcing the strict additivity of global momentum conservation, we arrive at a stark limitation: a purely classical mediator $E$ is fundamentally incapable of inducing any variation in the local momentum observable of the quantum mass $S$. Mathematically, the local momentum remains completely frozen:
\begin{equation}\langle P_S\rangle_{t} = \langle P_S\rangle_{t+\Delta t},\end{equation}
where $\langle P_S\rangle_t= \text{Tr}(P_S\varrho)$ and $\langle P_S\rangle_{t+\Delta t}= \text{Tr}(P_S\varrho^\prime )$. At first glance, this conclusion appears to contradict classical intuition, where standard Newtonian gravity routinely alters particle momentum. However, this apparent paradox dissolves when one scrutinizes the accounting of conserved quantities. In classical mechanics, momentum conservation is typically imposed directly between mass $S$ and mass $M$ via instantaneous action-at-a-distance, entirely bypassing the field as a momentum-carrying entity. In our framework, the field $E$ is explicitly concretized as a local interacting system. When local momentum conservation is rigorously enforced at the $S-E$ and $E-M$ interfaces within the confines of a classical channel, the dynamical transfer of momentum to the quantum mass is strictly forbidden.

To formally understand why a quantum gravitational field $E$ bypasses the restrictions of our no-go theorem, we transition to the Heisenberg picture to examine the exact continuous evolution of the local momentum observable $\langle P_S \rangle$. In the fully quantum scenario, the composite system $S \otimes E \otimes M$ undergoes global unitary evolution driven by the Hamiltonian $H_{SEM} = H_S + H_E + H_M + H_{SE} + H_{EM}$.The strict conservation of total momentum is encoded in the exact algebraic condition $\langle [P_S + P_E + P_M, H_{SEM}] \rangle = 0$. However, unlike a classical control map restricted to a single, non-fluctuating basis, the quantum interaction Hamiltonian $H_{SE}$ acts jointly across the tensor product Hilbert space. The dynamical evolution of the local momentum is strictly governed by the equation of motion:
\begin{equation}
    \frac{d\langle P_S \rangle}{dt} = -i \langle [P_S, H_{SEM}] \rangle.
\end{equation}
Because the mediator $E$ is a quantum field supporting non-commuting observables, the local momentum operator $P_S$ generally does not commute with the interaction term $H_{SE}$. Evaluating this commutator explicitly yields $[P_S, H_{SEM}] \neq 0$, which instantaneously dictates:$$\frac{d\langle P_S \rangle}{dt} \neq 0$$The change in the quantum particle's momentum is exactly compensated by the field and the massive object, such that $\frac{d}{dt}\langle P_E + P_M \rangle = -\frac{d\langle P_S \rangle}{dt}$. 
Compared to the classical gravitational field,  the quantum one can induce changes in the momentum of a quantum system for both $S$ and $M$. Consequently, assuming the proposed hybrid dynamics, if a particle is observed to change momentum after being subjected to a gravitational field, then the gravitational field should be quantum.


Let us consider a macroscopic source mass $M$, such as the Earth or a metal sphere, which can be effectively described as quantum matter approaching the classical limit (see Fig.~\ref{fig2}). Within our proposed framework, provided that global momentum is strictly conserved, any observable momentum transfer to the quantum mass $S$ mediated via the gravitational field constitutes sufficient evidence for the non-classicality of gravity. Remarkably, this paradigm recontextualizes several existing empirical observations. Phenomena such as the free fall of a quantum particle, or the spatial displacement driven by the gravitational field of a macroscopic sphere~\cite{GravAB}, fundamentally rely on this dynamical momentum exchange. Thus, under the stringent constraints of our no-go theorem, these standard observations already serve as compelling signatures of the quantumness of gravity.

\section{Other conservation laws}


Although our preceding arguments focused on momentum conservation, the underlying no-go theorem applies universally to any additive conserved quantity, including energy. When explicitly modeling the gravitational field as an independent interacting physical mediator, the local energy observable of a quantum particle corresponds strictly to its free energy (e.g., its kinetic energy), fundamentally distinct from any interaction or potential energy shared with the field. Dictated by our established theoretical constraints, if the strict global energy of the hybrid system is conserved, a classical gravitational field is completely prohibited from altering the particle's local free energy. Consequently, within the confines of this hybrid dynamics framework, the ubiquitous phenomenon of free fall—where a particle manifestly gains kinetic energy—serves as a stark violation of classical mediator restrictions, offering operational witness of the non-classicality of gravity.

\section{Conclusions}



In summary, we have established a general information-theoretic framework for hybrid quantum-classical dynamics (see Figure \ref{fig1}), fundamentally constrained by the classical system's inability to generate quantum correlations. Within this operational regime, we derive a rigorous no-go theorem: under the assumption of any additive conservation law, a classical mediator is strictly prohibited from altering the local observables of a quantum system. Applying this theorem to gravitational interactions unveils a profound tension between macroscopic conservation principles and strictly classical models of gravity. Specifically, the ubiquitous observation of dynamical momentum or energy transfer—such as the free fall of a quantum mass— suggests that the gravitational field must possess non-classical degrees of freedom to remain consistent with strict conservation laws.

Beyond gravity, our framework highlights how expectation values act as a universal bridge, reconciling disparate mathematical structures across physics. While derived in a non-relativistic context, the foundational nature of conservation laws implies broad validity for these limits. By linking rigorous dynamical constraints with everyday empirical observations, this work may open a new, operational pathway toward understanding the non-classical structure of spacetime.
\acknowledgments
T. F. thanks the support of QICI of the University of Hong Kong, and is grateful to Tian Zhang for helpful discussions. C. M. and V.V. thank the Gordon and Betty Moore Foundation. 	This publication was also made possible through the support of the ID 61466 grant from the John Templeton Foundation, as part of the The Quantum Information Structure of Spacetime (QISS) Project (qiss.fr). The opinions expressed in this publication are those of the Authors and do not necessarily reflect the views of the John Templeton Foundation.


\begin{widetext}
    
\section{Appendix}	

\subsection{Observables in general quantum-classical hybrid systems}
	
Different physical theories, such as classical mechanics and quantum mechanics, may exhibit incompatible mathematical structures. However, they share a fundamental operational feature: the ability to describe physical systems using expectation values. In terms of expectation values, both classical and quantum theories yield real numbers, and the validity of a theory is determined by whether its predicted expectations match empirical observations. Therefore, expectation values of observables (dynamical variables) serve as a universal bridge connecting different theoretical frameworks. From an operational perspective, a general description of the expectation value of an observable is strictly required.
	
In the physical world, empirical access is limited to the expectation values of observables. These expectation values are jointly determined by the physical state and the observable itself. We can succinctly formalize this mapping as:
\begin{equation}
	\langle \text{Observable} \rangle \equiv \{\text{State, Observable}\}.
\end{equation}
In quantum theory, the expectation value of an observable $O$ evaluates to:
\begin{equation}
	\langle O\rangle = \{\varrho, O\} = \text{Tr}(\varrho {O}),
\end{equation}
where $\varrho$ denotes the quantum density matrix. For a classical system, if it is in a mixture of several distinct pure states $\text{State}_i$ with probability $p_i$, the mixed classical state is denoted as $\wp_{\mathcal{C}} = \sum_{i}p_i\text{State}_i$. Using the generalized language defined above, the expectation value of a classical system is linearly described as:
\begin{equation}
	\{ \wp_{\mathcal{C}}, O\} \equiv \sum_{i} p_i \{ \text{State}_i, O\}.
\end{equation}
Crucially, if $\text{State}_i$ represents a state with a definitive scalar value $O_i$ for the given observable, this mapping exactly recovers the standard statistical expectation $\sum_{i}p_i O_i$. Thus, even without assuming a specific underlying mechanism for the classical sector, one can universally use this abstraction to represent the expectation value. 
	
Let us now analyze how hybrid systems are described within this framework. A hybrid classical-quantum state is algebraically represented via the direct sum $\mathcal{C}\oplus\mathcal{Q}$. For an additive global observable, we intuitively denote it as $O = O_\mathcal{C} \oplus O_\mathcal{Q}$. The expectation value of this additive observable evaluated on the hybrid state is strictly given by:
\begin{equation}
	\{\text{Hybrid State}, O\} = \{\text{Hybrid State}, O_\mathcal{C} \oplus O_\mathcal{Q}\}.
\end{equation}
Because the quantum and classical sectors of a direct-sum hybrid system are mathematically prohibited from generating quantum correlations (entanglement), the most general correlation permissible is purely classical. Therefore, the global expectation value linearly decomposes into local contributions:
\begin{equation}
	\{\text{Hybrid State}, O_\mathcal{C} \oplus O_\mathcal{Q}\} = \langle O_\mathcal{C}\rangle + \langle O_\mathcal{Q} \rangle,
\end{equation}
where
\begin{equation}
	\begin{split}
		\langle O_{\mathcal{C}}\rangle &= \{ \text{Reduced Classical State}, O_\mathcal{C} \}, \\
		\langle O_{\mathcal{Q}}\rangle &= \{ \text{Reduced Quantum State}, O_\mathcal{Q} \}.
	\end{split}
\end{equation}
	
Suppose the initial state of the hybrid system $\chi_{\mathcal{QC}}$ possesses no classical correlation, denoted as $\chi_{\mathcal{QC}} = \wp_{\mathcal{C}} \oplus \varrho$. Then,
\begin{equation}
	\langle O_\mathcal{C} \oplus O_\mathcal{Q} \rangle = \{\wp_{\mathcal{C}} \oplus \varrho, O_\mathcal{C} \oplus O_\mathcal{Q}\} = \{ \wp_{\mathcal{C}},O_\mathcal{C}\} + \{ \varrho, O_\mathcal{Q}\} = \langle O_\mathcal{C} \rangle + \langle O_\mathcal{Q} \rangle.
\end{equation}
For a generally correlated initial hybrid state $\chi_{\mathcal{QC}} = \sum_{j}q_j \wp_{\mathcal{C}j} \oplus \varrho_j$, where $\wp_{\mathcal{C}j} = \sum_{i}p^{(j)}_i\text{State}^{(j)}_i$, one generally has:
\begin{equation}
	\{\sum_{j}q_j \wp_{\mathcal{C}j} \oplus \varrho_j, O_\mathcal{C} \oplus O_\mathcal{Q}\} = \sum_{j}q_j \{ \wp_{\mathcal{C}j}, O_\mathcal{C}\} + \sum_{j}q_j\text{Tr}( \varrho_j O_{\mathcal{Q}}) = \langle O_\mathcal{C} \rangle + \langle O_\mathcal{Q} \rangle.
\end{equation}
Under general hybrid dynamics, the evolution from an initially uncorrelated state naturally generates classical correlations. The corresponding evolution of the expectation value over an infinitesimal slice is:
\begin{equation}
	\{\wp_{\mathcal{C}} \oplus \varrho, O_\mathcal{C} \oplus O_\mathcal{Q}\} \xrightarrow{V} \{V_\mathcal{C}\cdot \wp_{\mathcal{C}} \oplus \varrho, O_\mathcal{C} \oplus O_\mathcal{Q}\} \xrightarrow{\Lambda} \{ \sum_{i} q_i\text{State}_i \oplus \Lambda_i \cdot \varrho, O_\mathcal{C} \oplus O_\mathcal{Q}\} = \langle O_\mathcal{C} \rangle_{t+\delta t} + \langle O_\mathcal{Q} \rangle_{t+\delta t},
	\label{evo}
\end{equation}
where $V_\mathcal{C}$ and $\Lambda_i$ represent the local classical map and the classical-to-quantum control map, respectively. The variation of local observables $\langle O_\mathcal{C} \rangle$ and $\langle O_\mathcal{Q} \rangle$ serves as an exact indicator of dynamical back-action between the classical and quantum sectors.

     \subsection{Multiple-time evolution of hybrid systems}
	
	In the main text, we have analyzed a general dynamics of the hybrid system for an infinitesimal time interval. It is shown that, under the conservation law of an additive observable $O$ (i.e., energy or momentum), any quantum-classical hybrid dynamics is fundamentally incapable of varying the local observables (corresponding to the conserved quantity) in either subsystem.
	
	Here, we analyze in more detail the dynamics and properties of the hybrid system in terms of its continuous evolution over multiple time intervals. Note that any local free operation on a quantum system can be incorporated into the control channel $\Lambda$, and local operations inherently do not change the conserved quantity $O_{\mathcal{Q/C}}$. Thus, without loss of generality, we make use of Fig. \ref{fig3} to investigate the sequential evolution of the hybrid system from $t_0$ to $t_{2n}$. 
	
	Similar to the analysis in the main text, the composed state of the hybrid system at an arbitrary even time step $t_{2m}$ is denoted as:
	\begin{equation}
		\chi_{\mathcal{QC}}(t_{2m}) = \sum_{i}q^{(2m)}_i \wp^{(2m)}_{\mathcal{C}i} \oplus \varrho^{(2m)}_i,
	\end{equation}
	where $q^{(2m)}_i$ characterizes the classical correlation between the quantum and classical sectors. It is worth noting that odd time moments correspond to $V$ operations locally acting on the classical system, while even ones correspond to $\Lambda$ operations acting as classical-to-quantum control channels, as shown in Fig. \ref{fig3}. Therefore, from an even to an odd step, the quantum state is strictly unperturbed, yielding $\varrho_i^{(2m-2)} = \varrho_i^{(2m-1)}$.
	
	Suppose the classical state at $t_{2m-1}$ is expanded in its basis as $\wp^{(2m-1)}_{\mathcal{C}i} = \sum_{j}p^{(2m-1)}_{ij} \text{State}_j$. The state of the hybrid system evolving from $t_{2m-1}$ (odd) to $t_{2m}$ (even) is given as:
	\begin{equation}
		\chi_{\mathcal{QC}}(t_{2m-1}) = \sum_{i}q^{(2m-1)}_i \wp^{(2m-1)}_{\mathcal{C}i} \oplus \varrho^{(2m-1)}_i \rightarrow 	\chi_{\mathcal{QC}}(t_{2m}) = \sum_{i,j} q^{(2m-1)}_i p^{(2m-1)}_{ij} \text{State}_j \oplus \Lambda_j^{(2m)} \cdot \varrho^{(2m-1)}_i.
	\end{equation}	
	That is, by re-indexing the state at $t_{2m}$, we have:
	\begin{equation}
		\chi_{\mathcal{QC}}(t_{2m}) = \sum_{k}q^{(2m)}_k \wp^{(2m)}_{\mathcal{C}k} \oplus \varrho^{(2m)}_k = \sum_{i,j} q^{(2m-1)}_i p^{(2m-1)}_{ij} \text{State}_j \oplus \Lambda_j^{(2m)} \cdot \varrho^{(2m-1)}_i. \label{2m-1to2m}
	\end{equation}
	
	Similarly, one can analyze the nature of the dynamics from $t_{2m-2}$ (even) to $t_{2m-1}$ (odd), where only the local classical operation $V^{(2m-1)}$ acts:
	\begin{equation}
		\chi_{\mathcal{QC}}(t_{2m-2}) = \sum_{i}q^{(2m-2)}_i \wp^{(2m-2)}_{\mathcal{C}i} \oplus \varrho^{(2m-2)}_i \rightarrow \chi_{\mathcal{QC}}(t_{2m-1}) = \sum_{i}q^{(2m-2)}_i \left( V^{(2m-1)} \cdot \wp^{(2m-2)}_{\mathcal{C}i} \right) \oplus \varrho^{(2m-2)}_i.
	\end{equation}

	\begin{figure}[tb]
		\center
		\includegraphics[scale=0.43]{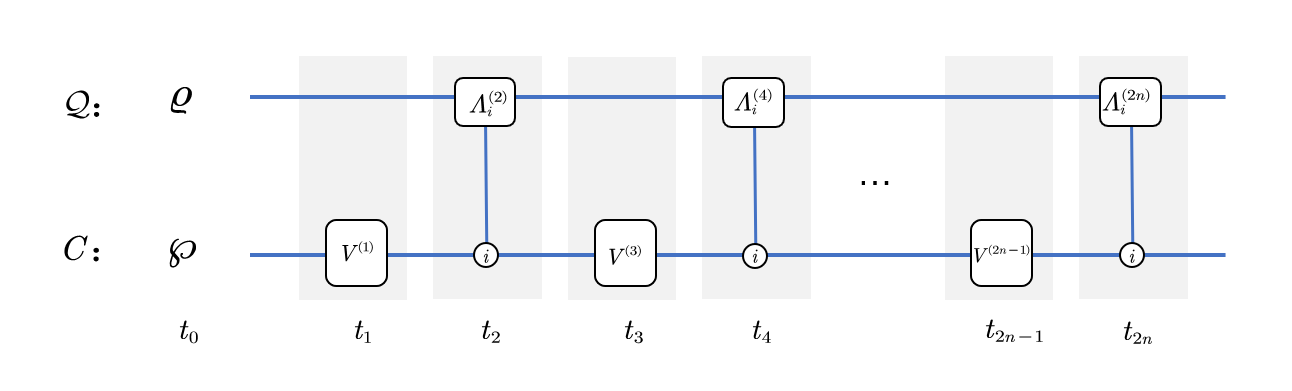}
		\caption{An effective quantum channel decomposition of quantum-classical dynamics for multi-time evolution. Each time interval is infinitesimal, $t_i - t_{i+1} = \delta t$. The dynamics of a hybrid system consist of these sequential infinitesimal slices of evolution. For simplicity, local free operations on the quantum system are incorporated into $\Lambda$, as they do not alter the conserved quantity $O_{\mathcal{Q/C}}$.}
		\label{fig3}
	\end{figure}
	
	Now let us assume there is a globally conserved additive observable $O_\mathcal{C} \oplus O_\mathcal{Q}$. The expectation values of this observable at time $t_{2m-1}$ and $t_{2m}$ are given respectively by:
	\begin{equation}
	\begin{split}
		\langle O_\mathcal{C} \oplus O_\mathcal{Q} \rangle_{t_{2m-1}} &= \{\chi_{\mathcal{QC}}(t_{2m-1}), O_\mathcal{C}\} + \{\chi_{\mathcal{QC}}(t_{2m-1}), O_\mathcal{Q}\} \\
		&= \sum_{i}q^{(2m-1)}_i \{ \wp^{(2m-1)}_{\mathcal{C}i}, O_\mathcal{C} \} + \sum_{i}q^{(2m-1)}_i \text{Tr}\left( \varrho^{(2m-1)}_i O_{\mathcal{Q}} \right),
	\end{split} \label{t2m-1}
	\end{equation}	
	and 
	\begin{equation}
		\langle O_\mathcal{C} \oplus O_\mathcal{Q} \rangle_{t_{2m}} = \{\chi_{\mathcal{QC}}(t_{2m}), O_\mathcal{C}\} + \{\chi_{\mathcal{QC}}(t_{2m}), O_\mathcal{Q}\} = \sum_{k}q^{(2m)}_k \{ \wp^{(2m)}_{\mathcal{C}k}, O_\mathcal{C} \} + \sum_{k}q^{(2m)}_k \text{Tr}\left( \varrho^{(2m)}_k O_{\mathcal{Q}} \right). \label{2m}
	\end{equation}
	
	Substituting Eq. (\ref{2m-1to2m}) into the classical part of Eq. (\ref{2m}), it is straightforward to verify that the classical expectation value remains entirely unchanged during the control operation, i.e., $\{\chi_{\mathcal{QC}}(t_{2m-1}), O_\mathcal{C}\} = \{\chi_{\mathcal{QC}}(t_{2m}), O_\mathcal{C}\}$. Thus, $\langle O_\mathcal{C} \oplus O_\mathcal{Q} \rangle_{t_{2m}}$ can be rewritten to explicitly match the classical part of $t_{2m-1}$:
	\begin{equation}
		\langle O_\mathcal{C} \oplus O_\mathcal{Q} \rangle_{t_{2m}} = \sum_{i}q^{(2m-1)}_i \{ \wp^{(2m-1)}_{\mathcal{C}i}, O_\mathcal{C} \} + \sum_{k}q^{(2m)}_k \text{Tr}\left( \varrho^{(2m)}_k O_{\mathcal{Q}} \right).
	\end{equation}
	
	Since the global observable is conserved, $\langle O_\mathcal{C} \oplus O_\mathcal{Q} \rangle_{t_{2m-1}} = \langle O_\mathcal{C} \oplus O_\mathcal{Q} \rangle_{t_{2m}}$. Subtracting the two equations yields:
	\begin{equation}
		\langle O_\mathcal{C} \oplus O_\mathcal{Q} \rangle_{t_{2m-1}} - \langle O_\mathcal{C} \oplus O_\mathcal{Q} \rangle_{t_{2m}} = \sum_{i}q^{(2m-1)}_i \text{Tr}\left( \varrho^{(2m-1)}_i O_{\mathcal{Q}} \right) - \sum_{k}q^{(2m)}_k \text{Tr}\left( \varrho^{(2m)}_k O_{\mathcal{Q}} \right) = 0.
	\end{equation}
	In other words, the local observable $O_{\mathcal{Q}}$ is strictly conserved from the odd to the even time moments:
	\begin{equation}
		\langle O_\mathcal{Q} \rangle_{t_{2m-1}} = \langle O_\mathcal{Q} \rangle_{t_{2m}}.
	\end{equation}	
	
	Similarly, from $t_{2m-2}$ to $t_{2m-1}$, the state is rewritten as:
	\begin{equation}
		\chi_{\mathcal{QC}}(t_{2m-1}) = \sum_{i}q^{(2m-1)}_i \wp^{(2m-1)}_{\mathcal{C}i} \oplus \varrho^{(2m-1)}_i = \sum_{i}q^{(2m-2)}_i \left( V^{(2m-1)} \cdot \wp^{(2m-2)}_{\mathcal{C}i} \right) \oplus \varrho^{(2m-2)}_i.
	\end{equation}
	Immediately, the observable expectation values for $t_{2m-2}$ and $t_{2m-1}$ are:
	\begin{equation}
		\begin{split}
			\langle O_\mathcal{C} \oplus O_\mathcal{Q} \rangle_{t_{2m-2}} &= \sum_{i}q^{(2m-2)}_i \{ \wp^{(2m-2)}_{\mathcal{C}i}, O_\mathcal{C} \} + \sum_{i}q^{(2m-2)}_i \text{Tr}\left( \varrho^{(2m-2)}_i O_{\mathcal{Q}} \right), \\
			\langle O_\mathcal{C} \oplus O_\mathcal{Q} \rangle_{t_{2m-1}} &= \sum_{i}q^{(2m-2)}_i \{ V^{(2m-1)} \cdot \wp^{(2m-2)}_{\mathcal{C}i}, O_\mathcal{C} \} + \sum_{i}q^{(2m-2)}_i \text{Tr}\left( \varrho^{(2m-2)}_i O_{\mathcal{Q}} \right).
		\end{split}
	\end{equation}	
	
	Obviously, from $t_{2m-2}$ to $t_{2m-1}$, there is no action on the quantum sector, leading directly to $\{\chi_{\mathcal{QC}}(t_{2m-2}), O_\mathcal{Q}\} = \{\chi_{\mathcal{QC}}(t_{2m-1}), O_\mathcal{Q}\}$. Namely:
	\begin{equation}
		\langle O_\mathcal{Q} \rangle_{t_{2m-2}} = \langle O_\mathcal{Q} \rangle_{t_{2m-1}}.
	\end{equation}	
	
	Thus, for a conserved global additive observable $O_\mathcal{C} \oplus O_\mathcal{Q}$, at any arbitrary time steps $t_m$ and $t_n$, we must have:
	\begin{equation}
		\langle O_\mathcal{Q} \rangle_{t_{m}} = \langle O_\mathcal{Q} \rangle_{t_{n}}, \quad \langle O_\mathcal{C} \rangle_{t_{m}} = \langle O_\mathcal{C} \rangle_{t_{n}}.
	\end{equation}	
	This mathematically guarantees that the expectation value of the local observable $O_{\mathcal{Q/C}}$ of either subsystem is an absolute conserved quantity under quantum-classical hybrid dynamics. Therefore, our no-go theorem applies rigorously to the continuous evolution of hybrid systems over any time scale.   

\end{widetext}


\begin{thebibliography}{99}
\bibitem{Reasons}
		If different orthogonal states of the quantum system control different operations of the classical system, then the superposition of these orthogonal states will lead to the transmission of quantum information to the classical system's thereby creating quantum correlations, which contradicts the definition of classicality.
\bibitem{statement}
		It is similar to the argument of conservation of momentum, except that the conserved quantity is replaced by energy in the argument. 
\end{thebibliography}
\end{document}